\documentclass{article}

\usepackage{subfigure}
\usepackage{graphicx}
\usepackage[numbers,sort&compress]{natbib}
\bibpunct[, ]{[}{]}{,}{n}{,}{,}

\begin{document}


\title{Using clustering of rankings to explain brand preferences with personality and socio-demographic variables}

\author{
Daniel M\"{u}llensiefen\footnote{Department of Psychology, Goldsmiths, University of London, UK}, Christian Hennig\footnote{Department of Statistical Science, University College London, UK, Email: c.hennig@ucl.ac.uk}, Hedie Howells\footnote{Department of Psychology, Goldsmiths, University of London, UK}}

\maketitle

\begin{abstract}
The primary aim of market segmentation is to identify relevant groups of consumers that can be addressed efficiently by marketing or advertising campaigns. This paper addresses the issue whether consumer groups can be identified from background variables that are not brand-related and how much personality vs. socio-demographic variables contribute to the identification of consumer clusters. This is done by clustering aggregated preferences for 25 brands across 5 different product categories, and by relating socio-demographic and personality variables to the clusters using logistic regression and random forests over a range of different numbers of clusters. Results indicate that some personality variables contribute significantly to the identification of consumer groups in one sample. However, these results were not replicated on a second sample that was more heterogeneous in terms of socio-demographic characteristics and not representative of the brands’ target audience. 
~\\~\\
{\it This preprint is currently submitted to the Journal of Applied Statistics.}
\end{abstract}


\section{Introduction}
The primary aim of market segmentation is to identify relevant groups of consumers that can be addressed efficiently by marketing or advertising campaigns. Here we address the issue whether consumer groups can be identified from background variables that are not brand-related, and particularly how much personality vs. socio-demographic variables contribute to the identification of consumer clusters. See Section \ref{smarket} for an overview of the relevant literature. 

We suggest a statistical approach based on clustering to generate consumer profiles and evaluate the association of socio-demographic and psychological variables with clusters of consumers. However, we do not make any assumptions about a ``true'' or ``natural'' number of consumer clusters, but rather to try to identify ``constructive'', pragmatic clusters (\cite{hennig2015}). In fact, we start from the presumption that the underlying multivariate distribution of brand preferences is continuous and use clustering as a tool to segment this continuous distribution into meaningful groups of individuals who differ in their brand preferences and therefore can be targeted differently with marketing communication. Here we take into consideration that in most practical applications the number of different target groups or market segments is determined by practical constraints dictated largely by logistic complexities and production costs that increase with the number of clusters, i.e. target groups that need to be addressed separately. Thus, we follow a profiling approach where we measure the impact of socio-demographic and psychological variables as a function of the number of clusters. For the marketing practitioner this translates to the question: ``For the number of different market segments I can target, how important are psychological vs demographic variables to describe the different clusters of consumers, and which specific variables are most important?''

We test the approach in studies using two different samples that are similar in size but differ in their socio-demographic characteristics. The first sample is drawn from a population of young urban consumers living in the London area that can be considered demographic target audience for the product categories and brands that we selected in this study based on data from the UK TGI database. The second sample is has a wider spread on all socio-demographic variables and is drawn from a panel representing Scottish adult consumers of all ages and social strata; i.e. it is much more diverse than the first sample and cannot be considered to represent as a whole the demographic target audience. Thus, the two studies represent different scenarios: The first study resembles a market research project where prior knowledge about a consumer target group for specific brands and products is used to narrow down the selection of a panel and the second study is closer to academic studies where broader (but at the same time often larger) samples are drawn from existing panels or databases representing the general adult population of consumers. 

In both studies we assess consumer preferences for five different brands drawn from each of five different product categories serve as the primary outcome data. Participants are clustered according to their ranked brand preferences. Cluster solutions from 2 up to 10 consumer clusters are computed. In a second step, each cluster solution is then assessed in terms of how well socio-demographic and personality variables contribute to predicting cluster membership of individual consumers.

Our primary focus is on  the contribution of the personality 
variables when the number of consumer clusters changes. We expect that the personality traits gain importance for describing the differences between clusters of consumers as the number of clusters increases, i.e. for finer grained market segmentations. We also expect to observe a stronger contribution of personality variables in Study 1 (target group sample) than in Study 2 (general consumer sample). By way of the same analysis we will also be able to test the assertion that neither demographic nor personality variables are significantly associated with consumer clusters (e.g. \cite{sharp2010brands}). In order to address these issues, we use logistic regression and random forests for explaining the clusters from the explanatory variables. 

Section \ref{smarket} summarises the relevant literature on market segmentation.
Section \ref{sstudies} gives some details about the two empirical studies. 
Section \ref{sstat} presents that statistical methodology that was applied here.
Section \ref{sresults} summarises the results. Section \ref{sdiscussion} concludes the paper with a discussion.

\section{Relevant literature on market segmentation and personality variables} \label{smarket}
Market segmentation is based on the assumption that an individual product or brand will not appeal equally to the entire population and that marketing budgets can be used most effectively when marketing communications are tailored to the characteristics of specific consumer groups who are most likely to purchase a product or engage with a brand (\cite{lilien2002marketing,peter2010consumer}, p. 363) and thus saving considerable resources in comparison to a scattered, ‘shotgun’ approach (\cite{fitzgerald2005products}, p. 512). In addition, market segmentation can also aim to segment consumers by media consumption or by their value to the brand (e.g. loyal vs. occasional consumers of the brand). But overall the primary aim of market segmentation is the identification of target consumer groups (\cite{blamires2000segmentation}, p. 499) and to inform marketers how, when, and where to advertise, as well as indicating ‘who’ comprises the target market.
 
While developing new ideas for brand and product communications, advertising creatives often employ simplified notions of ‘prototypical consumers’ that are the intended audience(s) for specific communications. Similarly, marketing practitioners make use of insights into the behaviour of target consumer groups to maximise the effectiveness of marketing campaigns, for example for media planning and forecasting the effects of campaigns on sales and revenue.   

For most practical applications in advertising and consumer marketing, market segmentations are performed on the basis on basic demographic information, most commonly and with little variation over the past decades (\cite{michman1991lifestyle,sandy2013predicting}), mainly age, gender, household income, and ethnicity. In addition, market segmentation is often based on standard classification schemes for social class (e.g. the 6 occupation-based groups of the NRS scheme as maintained by the Market research Society in the UK) or on neighbourhood area (e.g. the ACORN or Mosaic schemes in the UK). Larger market segmentation schemes, such as the PRIZM NE scheme in the US, combine demographic as well as geographic and neighbourhood with lifestyle data to yield very fine-grained consumer classifications. Information from these standard industry classification schemes is readily available together with product preferences through large survey panels such as the ones provided by MINTEL, the Target Group Index (TGI) or Market Assessment.  

In contrast, the term ``psychographics'' (\cite{furnham1992consumer,demby1994psychographics}) describes the collection of data on variables that reflect consumer personality, attitudes, personal values (\cite{kahle1983social,mitchell1983nine}), life style (\cite{michman1991lifestyle}) and other psychological constructs in order to identify and describe subpopulations of consumers and ultimately to inform marketing processes (see a recent summary account on the use of psychographics in market segmentation in \cite{sandy2013predicting}). Early empirical studies testing the psychographic approach only found limited evidence for the predictive value of personality variables impacting on consumer behaviour (\cite{brody1968personality,thumin1968consumer,lehmann1971television,kahle1986alternative}).  Since then psychographic approaches to market segmentation have been repeatedly criticised for being ineffective and only explaining a small amount of variance in consumer choices (\cite{wells1975psychographics,novak1990comparing}). However, over the last two decades personality psychology has seen great advancements both in terms of psychometric developments, both through increasing the reliability and validity of the measurement of personality traits as well as in terms of the applications of personality measurement ranging from behavioural genetics (\cite{plomin1990behavioral} to core applications in human resources selection and work place contexts (\cite{salgado1997five}) and important life outcomes in general (\cite{roberts2007power}). In addition, the Big Five model of personality (\cite{costa1992neo,mccrae1999five}) - comprising Extraversion, Neuroticism, Agreeableness, Conscientiousness, and Openness to Experience - has been established as a quasi-standard in the conceptualisation and empirical measurement of personality (\cite{john2008paradigm}), which enables broad comparisons of research findings across different studies and fields. As a consequence of these recent developments in personality psychology, several academics (e.g. \cite{baumgartner2002toward,miller2009spent}) have argued that - with a few exceptions (e.g, \cite{choo2004type,mulyanegara2009big}) -  personality variables have received their due credit yet in marketing research and practice and that the consideration of personality traits can play a crucial role in understanding consumer behaviour (\cite{ferguson2011personality}).
 
In academic as well as applied contexts market segmentation studies are typically performed either at the level of individual brands or the level of the product category. In these scenarios the research goal is to identify whether or not demographic variables or psychological traits co-vary with the choice of specific brands (\cite{hammond1996market}). However, there has been mixed support for the usefulness of demographic and psychological variables in market segmentation (\cite{novak1990comparing,kennedy2001competing,kennedy2001there,proc-anzmac-2006}, leading Sharp \cite{sharp2010brands} (p. 71) to suggest that predicting brand preferences from personal consumer variables is by and large impossible: ``The big discovery (of research on market segmentation) is that customer bases of brands in a category are very similar (\ldots), there isn’t a vanilla ice-cream buyer and a different type of person who buys strawberry -  there are just ice-cream buyers who sometimes buy vanilla and very occasionally buy strawberry''.
  
In a recent comparative study on data from a large sample, Sandy et al. \cite{sandy2013predicting} tested the relative contribution of demographic and personality factors in regression models for a large number of individual outcome variables. Outcome variables were each derived from specific statements on individual aspects of media consumption, political and societal views and product choices. In line with Sharp’s assertion, Sandy et al. found only very small effects, even when demographic and personality variables were combined in the same model – most of their models explained less than 10\% of the variance in the outcome variable. This is in line with the low effect sizes reported by \cite{novak1990comparing} where the median $R^2$ value for regression models using demographic variables was 0.04 and median $R^2$ values for models using List of Values and Values and Life Style psychographic variables were 0.011 and 0.026 respectively. In contrast, \cite{sandy2013predicting} found that personality variables from a Big Five inventory contributed about an equal amount to the regression models of their outcome variables of direct and indirect consumer behaviour.  In their discussion, \cite{sandy2013predicting} attribute the low effect sizes for both types of variables to the fact that their outcome variables only described a narrow aspect of direct and indirect consumer behaviour and they suggest to aggregate indicators of consumer behaviour in order to find larger effects. In addition, they suggest considering brand choices rather than choices for classes of products as primary outcome measures, assuming that personality (as well as demographic) variables may have a larger impact on brands choice. 

This study takes up these suggestions from \cite{sandy2013predicting} by starting from brand preference rankings within product categories and by using clustering methods to obtain aggregate consumer indicators. Then, using these methodological refinements, we assess (similar to \cite{sandy2013predicting}) the absolute and relative importance of personality and demographic variables for segmenting consumers into different groups. In addition, we assess the question of how the relative contribution of personality and demographic variables might change with the number of different consumer groups in a clustering solution. Hence, this study does not assume that the contribution of personality variables is constant across the number of market segments that a created during the segementation process but might depend on the complexity of the segmentation solution. 

It is worth noting that the term psychographics refers to a larger set of measures and scales that aim to capture consumer characteristics and behavior. Commonly, activities and interests, commercially relevant attitudes and behaviours as well as values and beliefs along with personality questionnaires are part of the psychographic toolbox in consumer research \cite{weinstein2004handbook}. However, in terms of practical research applications, psychographic information is usually much more difficult to obtain than demographic information because individuals have to be contacted individually and incentivized to self-report using non-standardized questionnaires. By contrast, demographic information at the household level can be obtained from databases such as the one maintained by PRIZM NE without the need to contact contact consumers directly. Thus, for most practical market segmentation applications, demographic information is already available for most Western countries (at least at the household-level), while psychographic  information is generally not available. However, personality constitutes an exception to this rule. The greater availability of personality data compared to other psychograhic measures is helped by the fact  that allmost all recent studies use the standard big five model of personality and hence regardless of the actual self-report personality inventory used, data and results can be compared across studies. In addition, recent studies have shown that it is not necessary to obtain personality data from consumers directly through self-report inventories but that personality information can be gathered indirectly from online data such as facebook likes \cite{Youyou27012015}, Twitter profiles \cite{QuerciaD2011Our}, musical preferences \cite{doi:10.1177/1948550616641473} or spending behavior \cite{doi:10.1177/0956797616635200}. Thus, unlike other psychographics measures, personality has become a layer of information that can be obtained from various sources, especially online information, and that has proven to be useful for predicting a broad variety of outcome measures, such as substance use, political attitudes, or purchase satisfaction \cite{Youyou27012015, doi:10.1177/0956797616635200}. Hence, this study focuses on the big five personality profile as the only psychograpohic measure to predict aggregate consumer preferences in addition to common demographic information.
    
Much of the traditional academic research on market segmentation has focused on predicting individual brand choices. However, it has been shown that consumer brand preferences can vary by context and that consumers can have multiple preferences within the same category (\cite{desarbo2008}), which can be one cause of the instability of preference models based on single brand choices. In line with the suggestion by Sandy and collaborators \cite{sandy2013predicting}, a more stable approach may result from aggregating brand preferences into broader consumer profiles across product categories, reflecting associations between multiple brands as opposed to a single preference. Aggregating consumer choices into consumer profiles or clusters often feeds into successful product recommendations (\cite{resnick1997recommender,ansari2000internet}) and prediction tasks. Thus, at least in online retailing, aggregating consumer choices has become a standard approach and been implemented and refined successfully by major online retailers and media services such as Amazon and Netflix. Similar approaches (e.g. \cite{dolnicar2010,mooi2011}) have been used very successfully with different types of survey data, e.g. from tourism research \cite{dolnicar2002} where the goal is to ``identify groups of tourists who share common characteristics'' and target them with a ``tailored marketing mix'' (\cite{dolnicar2012}).

\section{The studies}\label{sstudies}
\subsection{Methods}
For Study 1,
data was collected via a survey (paper as well as online) that was distributed to a sample of young adults in London, UK, in 2011-2012.

The aim of Study 2 was to assess to what degree the results from Study 1 hold true with data from a different sample which does not represent the brands’ demographic target audience. The data for this
was collected via an online survey that was distributed to a sample from the ScotPulse panel of adults living in Scotland, in 2012. The ScotPulse panel has around 12,000 members from all age and demographic groups and can be considered representative of the Scottish population. The panel is run commercially by STV.TV Ltd, Glasgow.

\subsection{Participants}
Study 1 had 343 participants (57.7\% women) with a mean age of 23.18 years (SD$=3.48$), which were recruited among young adults all currently residing within in the Greater London Area. 
Two thirds of the participants were from the target age group of up to 24 years while about 1/3 were between 25 and 34 years old. Participants were mainly living as singles (50\%), renting an accommodation (67\%), had achieved A-levels (44\%) or an undergraduate degree (41\%) as their highest level of education, and indicated a diverse range of income brackets for the main earner in their household (almost equal proportions for income brackets from $<$\pounds\ 15,000 to \pounds\ 50,000 - \pounds\ 75,000 per year).   

Study 2 had 355 participants (54.1\% women), who were recruited from among members of the ScotPulse panel living in urban as well as rural areas of Scotland. The age distribution was markedly different from the sample in Study 1 and included a larger proportion of older participants outside the age range for which we selected the 25 brands. 30.7\% of the participants were less than 25 years and 64.8\% were between 25 and 34 years old. 42.2\% lived in a 2-person household and for 29.0\% the yearly household income was between \pounds\ 15,000 and \pounds\ 30,000. For 27.3\% the household’s main earner was working in a professional or technical profession that requires at least university degree-level qualification. For 18.9\% the main earner worked in a non-managerial but non-manual job (e.g. office worker). 45.4\% of the participants had achieved an undergraduate degree as their highest level of education. As in Study 1, there were no missing values in the brand choice variables. 8 observations had missing values for one or more socio-demographic variables (one observation in Study 1).

\subsection{Materials}
Participants took an online survey that was distributed via chain-referral system in February to June 2012. The survey questionnaire was entirely anonymous and participants were not remunerated for their participation. The survey comprised the following measurement instruments.
\begin{description}
\item[Personality]
With a view on possible implementations in practical contexts we chose the Ten Item Personality Inventory (TIPI, \cite{gosling2003very}), which comprises only 10 brief statements that participants respond to on a 7-point agreement scale and has been shown to have high validity and reliability scores compared to other short-form Big Five personality inventories (\cite{furnham2008relationship}). Average scores for the five dimensions Extraversion, Agreeableness, Conscientiousness, emotional Stability, and Openness to Experience were derived for each participant.
\item[Socio-Demographic Variables]
In line with marketing practice and previous comparative literature (e.g. \cite{novak1990comparing,proc-anzmac-2006,sandy2013predicting}) we focused on the most widely used socio-demographic variables that are of relevance with the sample, namely age, gender, education, household main earner’s income, dwelling type and relationship status.
\item[Brand Choice]
Participants provided preference rankings for each of five brands from each of five different categories. The product categories were chosen to represent largely gender-neutral commodities with a high relevance to the participant target group (young adults living in an urban area). Brands in each category were chosen to be the five most frequently used brands in the age group of the 19-29 year olds according to data from the TGI GB survey for 2011. Categories and chosen brands were:
\begin{description}
\item[Smart phone:] iPhone, Nokia, Blackberry, Samsung, Sony Ericsson.
\item[Chocolate snack:] Dairy Milk, Galaxy Milk, Kit Kat, Maltesers, Creme egg.
\item[Clothing retailer:] Topshop/Topman, River Island, H\& M, Primark, George at ASDA.
\item[Coffee shop:] Starbucks, Café Nero, Costa Coffee, Department store’s own coffee shop, Local coffee shop.
\item[TV show:] X-Factor, Live at the Apollo, The Simpsons, Shameless, Harry Hill’s TV Burp.
\end{description}
\end{description}
In Study 2, participants took an online survey that was distributed randomly to the members of the ScotPulse panel. The survey questionnaire was entirely anonymous and participants were remunerated with a typical panel credit or their participation. Similar to the questionnaire instruments used with the London sample in Study 1, the survey in Study 2 used the TIPI as a personality inventory and questions on the socio-demographic variables age, gender, household main earner’s income, the work position of the household’s main earner  (similar to the ESeC classification of socio-economic status), the number of children living in the shared home and highest level of education achieved. Participants ranked the same 25 brands that were used in Study 1 according to their preferences. 

\section{Statistical methodology}\label{sstat}
\subsection{Cluster analysis and distances: methodology}
Market segmentation is done by clustering the brand choices. The resulting clusters are then used as response variable to be explained by the socio-demographic and personality variables. We are particularly interested in whether there is evidence that the personality variables have an impact on the brand choice clusters. The present section explains what was done. Section \ref{scdd} presents a thorough discussion of the choices involved in this approach.

The brand choices for a participant $i=1,\ldots,n$ can be represented as a 25-dimensional vector ${\bf r}_i=(r_{ijk})_{j=1,\ldots,5,k=1,\ldots,5}$, where $r_{ijk}$ is the rank given to brand $k$ in category $j$ by participant $i$, and for fixed $i, j$:
$\{r_{ijk}:\ k=1,\ldots,5\}=\{1,2,3,4,5\}$, where 1 denotes the first preference.
We use ``partitioning around medoids'' (PAM, \cite{kaufman1990}), which is a distance-based method for cluster analysis; for a given number of clusters $G$, PAM looks for the $G$ centroids in the dataset such that the sum of the distances of all objects to their closest centroids is minimised.  Rather than postulating that there is a single true number of clusters and trying to estimate it, we use all PAM solutions for $G=2,\ldots,10$ in order to monitor how the contribution of the socio-demographic and personality variables looks like over all the values for $G$ (we tentatively looked at $G>10$ but this did not bring any additional insight). 

A key issue for the definition of the distance between brand choices is that usually regarding brand preferences the higher ranks can be assumed to be more important than the medium and lower ranks; many consumers have favourite brands but would not differentiate much between several brands for which their preference is not so high, see also \cite{brentari2016}. For this reason we introduce a score function $s$ for the ranks, with $s(1)=1, s(2)=5, s(3)=7, s(4)=8, s(5)=9$. We then define the distance between two rankings as the ``Footrule''- ($L_1$-)distance between scores: 
\begin{equation}\label{e:sfootrule}
d({\bf r}_{i_1},{\bf r}_{i_2})=\sum_{j=1}^5\sum_{k=1}^5|s(r_{i_1jk})-s(r_{i_2jk})|. 
\end{equation}
We also ran analyses separately for each category, for which the definition of $d$ only requires a single sum over the 5 brands.

In order to assess the degree of ``clustering'' in the data (on which our methodology does not rely, although it may be interesting in its own right), we show 2-dimensional plots produced by classical Multidimensional Scaling (MDS, \cite{mardia1978}) as implemented in the R-function ``cmdscale'', and we compute two cluster validation indexes, namely the Average Silhouette Width (ASW) and the normalised version of Hubert's $\Gamma$ based on the Pearson correlation (P$\Gamma$, \cite{kaufman1990,halkidi2015}). The former assesses the quality of a clustering based on the contrast between within-cluster homogeneity and between-cluster separation with high values if there are clear gaps between the clusters. The latter formalises to what extent the distance is represented by the ``clustering-induced distance'', which is 0 between two observations in the same cluster and 1 between two objects in different clusters. This is particularly relevant here because the clusterings are used, even in absence of a true clear clustering of the data, to summarise the information in the brand preferences encoded by the distances.

\subsection{Cluster analysis and distances: discussion}\label{scdd}
The main aim of cluster analysis here is pragmatic. The information in the rankings is very complex, and the standard methods used in Section \ref{sexp} could not have been applied for explaining the full information in the rankings directly. Given the moderate sample sizes, some information reduction is required in order to apply any method for exploring how the explanatory variables affect the rankings. Forming clusters is our approach to summarize the ranking information. Occasionally, cluster analysis is used for reducing more complex information for use in explanatory methods (e.g., \cite{ammann2014}), but the specific requirements of such clusterings are rarely discussed. \cite{hennig2015} calls such a pragmatic clustering task ``constructive'', meaning that the aim is not to find true underlying ``real'' clusters, but rather to organize the data in a suitable way for the requirements of the specific application. 

In Section \ref{sexp}, the clusters are treated as a response for multinomial logit regression and random forests, and we are interested in whether explanatory variables have an impact on the brand rankings that are encoded through the clusters. Therefore it is important here that the clusters are homogeneous, i.e., that the within-cluster dissimilarity of rankings is low (otherwise it would be problematic to interpret the impact of the explanatory variables in terms of the rankings). We also want to represent the similarity structure well by the clustering, which is measured by P$\Gamma$. Another desirable feature of the clusters is that their sizes should be fairly uniform, so that the observations can be used in a balanced manner for predicting all clusters in Section \ref{sexp}. 

Separation of the clusters is less important (actually the MDS plots and also not shown higher dimensional MDS information make us believe that there are no strongly separated clusters in the data), and it is not desirable here to rely on the model assumptions required for postulating and fitting a probability model for the ranks. This also means that we do not use any definition of a ``true'' number of clusters $G$, for which reason we do not attempt to estimate $G$. Instead, we track results over various values for $G$.

Because there is also no particular reason why a hierarchy of clusters should be imposed, we decided to use the non-hierarchical PAM method that focuses on homogeneity of clusters rather than separation. Because of the triangle inequality, no within-cluster distance can be bigger in PAM than the sum of the two biggest distances of objects to the cluster centroid. For comparison, we ran Average Linkage and Complete Linkage hierarchical clustering, cutting the dendrograms at $G=2,\ldots,10$ clusters. This led to clearly worse values in all cases (both studies, all values of $G$) regarding the average within cluster distance and the uniformity of cluster sizes, and in almost all cases regarding ASW and P$\Gamma$.

The Footrule distance defined in (\ref{e:sfootrule}) is an $L_1$-distance on the vector of scores, and as such it fulfils the standard properties of a metric (identity, symmetry and the triangle inequality). We follow the philosophy of \cite{hennig2013find,hennig2015} here, according to which distance design is about formalisation of what counts as ``similar'' or ``distant'' in the given application, which cannot be estimated from the data alone but will always involve user input. The main feature that we wanted to achieve with the choice of the distance and the specific scores is to emphasise the first rank compared with the differences between the others, as we believe that consumers often have a ``favourite brand'', which often will have an impact on their buying behaviour, whereas we rather expect, at least for the lower ranks, that customers will normally construct a ranking only when asked, and that this ranking will have far weaker implications for their behavior. We quantified this in a subjective manner, but there is no objective alternative, because there is no information in the given data about this (one could imagine data from certain questionnaire questions or buying behavior that gives some information about how much more important the first or higher ranks are to a typical consumer, but such data was not available to us).
Note that the Footrule distance with the scores $s$ defined above is not rank-invariant (invariant against applying the same permutation to both rankings, \cite{marden1996analyzing}), because this is in direct conflict to emphasising the top ranks. It is label-invariant, i.e., invariant against permutation of the labels (listed brands) as rank differences are just summed up over brands.

The 25 score distances are aggregated in the Footrule distance in an $L_1$-manner, which seems appropriate to us because it gives the rank difference for every brand the same weight, whereas the Euclidean distance uses squares, which upweights brands with larger rank differences. 

 
There are a number of alternatives in the literature. \cite{brentari2016} base a distance on a generalisation of Spearman's rank correlation and the directly related Spearman's distance, which also gives higher weight to the better ranks, but in less pronounced and less direct manner (according to their definition, rank 3 is still more distant from rank 5 than rank 1 from rank 2); see also \cite{dancelli2013} for a comparison of various versions of Spearman's correlation. Ignoring the issue of giving higher weights to better ranks, there are some distances for ranking data other than Spearman's. \cite{marden1996analyzing} lists  for example Kendall's $\tau$- and the Footrule distance and compares the distances  in certain ways, most of which are not very relevant to clustering. We are not aware of strong arguments regarding the choice between Kendall's $\tau$-, Footrule and Spearman's $\rho$-distance relevant in our setup and stick to the Footrule distance because of the straightforward implementation of our nonstandard scoring and because $L_1$-aggregation seems to be most appropriate over different brands.

There is also some work on model-based clustering of rank data, see, e.g., \cite{croon1989,croon1993,murphy2003,gormley2006,gormley2008,gormley2009,jacques2014}. \cite{murphy2003} is distance-based and may yield good within-cluster homogeneity. As already stated, we are not convinced that assuming that there is an underlying ``true'' partition to be estimated is helpful in the given situation. Most of these methods do not readily generalize to nonstandard scorings and multivariate rankings, although the Rankcluster package in R \cite{jacques2014} analyses multivariate rankings. 
We applied this to our data but the resulting clusterings had extremely imbalanced cluster sizes, which in our situation is not desirable. Note that all comparisons between candidate methods referred to in this section were run ignoring the explanatory variables, in order to avoid biasing the results in Section \ref{sexp}.

We acknowledge that the model-based approach comes with its own advantages, which include an account of the choice process in some models for rank data \cite{luce1959,fligner1988} and a generic possibility to quantify uncertainty (although this is based on model assumptions that we are not willing to make); so there is certainly some potential for future work in this direction for similar applications; however, we think that the approach taken by us is superior for producing homogeneous clusters (regarding within-cluster distances and cluster sizes), which is a key issue for our pragmatic use of clustering for information summary. 

\subsection{Explaining the clusters from socio-demographic and personality variables} \label{sexp}
Once the clustering has been obtained, we use two different approaches to explain the clusters (which can be seen as simplifying proxies for the full ranking information) from the socio-demographic and personality variables. 

The first one is the multinomial logit regression (MLR) model (\cite{hosmer2000}). 
Let $y_i\in\{1,\ldots,G\}$ be the cluster to which the brand choices of participant $i$ belong, and $Y_i$ the corresponding random variable. Let ${\bf x}_i=(x_{ij}), j=1,\ldots,p$ be the vector of the values of explanatory variables for participant $i$. The MLR assumes that $Y_1,\ldots,Y_i$ are independent and 
\[
\log\left[\frac{P(Y_i=g)}{P(Y_i=1)}\right] =\beta_{g0}+\beta_{g1}x_{i1}+\ldots+\beta_{gp}x_{ip}
\]
for $g=2,\ldots,p$, with cluster 1 as reference category. We estimate this model  using the function ``multinom'' in the R-package ``nnet'' (\cite{MASS02}). Assume that the variables $j=1,\ldots,q$ are the socio-demographic variables and variables $j=q+1,\ldots,p$ are the personality variables. We are then particularly interested in the deviance test of the null hypothesis $\beta_{gj}=0$ for all $g=2,\ldots,G, j=q+1,\ldots,p$, meaning that none of the personality variable contributes significantly to explaining the brand choice clusters in the presence of the socio-demographic variables. The individual tests for all variables $j$ separately, testing $\beta_{gj}=0$ for all $g=2,\ldots,G$, are also considered, i.e., the null hypotheses that there is no contribution of variable $j$ beyond what can be explained by the other variables. Analyses here are conditional on the clusterings, which have been derived independently of the explanatory variables. No formal model selection has been applied, rather the presented model comprising all variables of interest without interactions is the biggest one that seems reasonable with the given numbers of observations. Therefore the standard theory of the MLR applies.

Because we do not fix the number of clusters, this leads to a large number of tests being performed. Statistical hypothesis tests and p-values have been criticised because in many studies including the present one the computation of multiple p-values and the potential of even more testing because of pre-processing decisions that could have been made differently (such as our scoring of ranks) make it too easy to find individual significant p-values (e.g., \cite{gelman2014}). We use p-values here in an exploratory manner for visualisation, highlighting particularly consistent (or inconsistent) significances over all or many numbers of clusters, without interpreting individual p-values according to the classical theory, which would implicitly assume that each of them was the only one we computed.

As a second approach we fitted random forests (\cite{breiman2001random}) as implemented in the R-package ``randomForest'') to the classification problem of predicting $Y_1,\ldots,Y_G$ from ${\bf x}_1,\ldots,{\bf x}_p$. The random forest is a powerful data mining technique and in contrast to the multinomial logistic regression model, random forests allow to model non-linear relationships and higher-order interactions between predictor variables. On the other hand, a lack of distribution theory does not allow to compute standard significance tests.

Random forests classify observations by majority vote based on many classification trees computed from nonparametric bootstrap samples. The influence of specific predictors in the presence of the other predictor (explanatory) variables by computing the mean decrease of classification accuracy for models that do not include a specific predictor variable. This and the overall classification error are computed as out-of-bag (OOB) error, i.e., for each observation only those trees are used for which the observation was not involved in growing the tree.

\section{Results}\label{sresults}
\begin{figure}[t]
\centering
\includegraphics[width=0.48\textwidth]{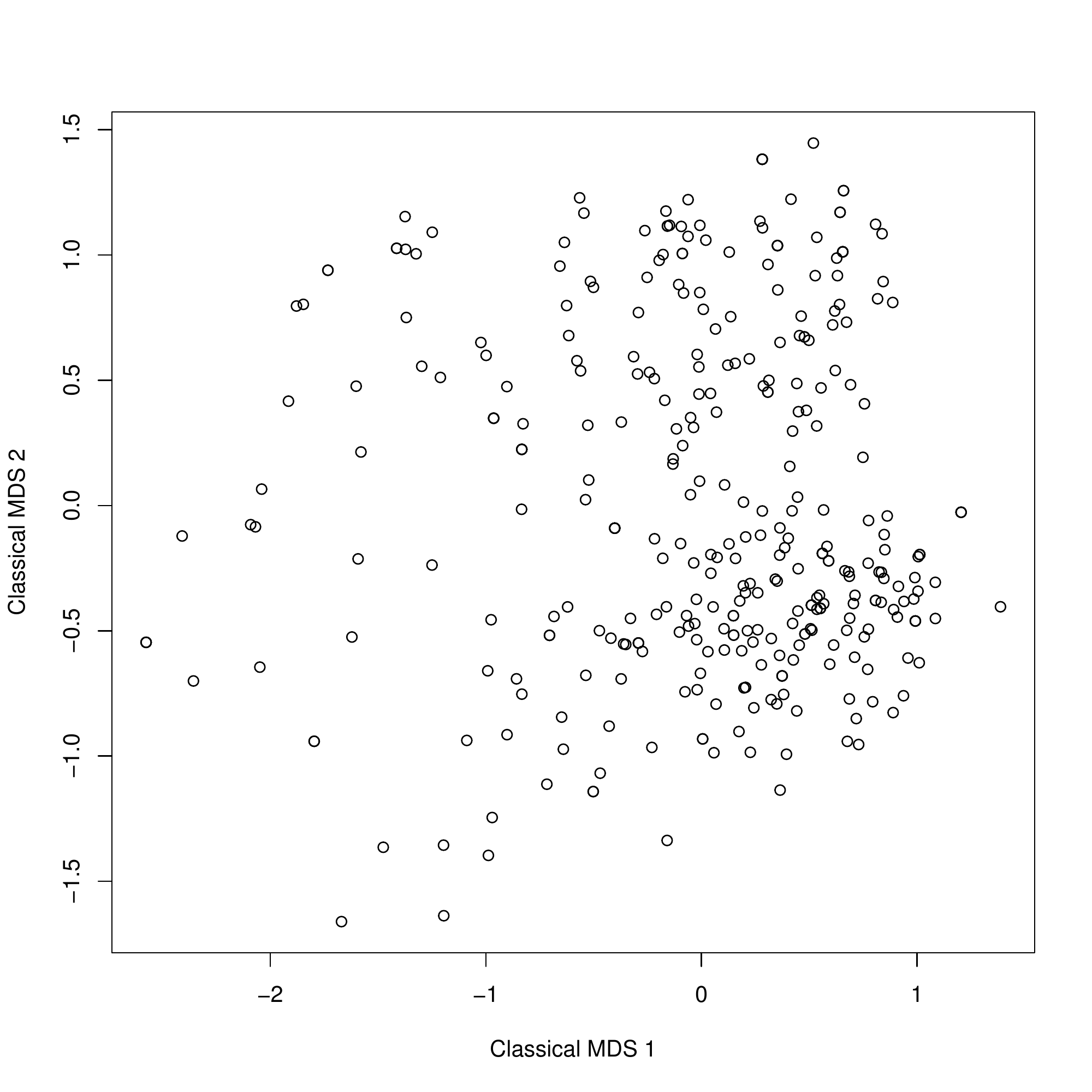}
\includegraphics[width=0.48\textwidth]{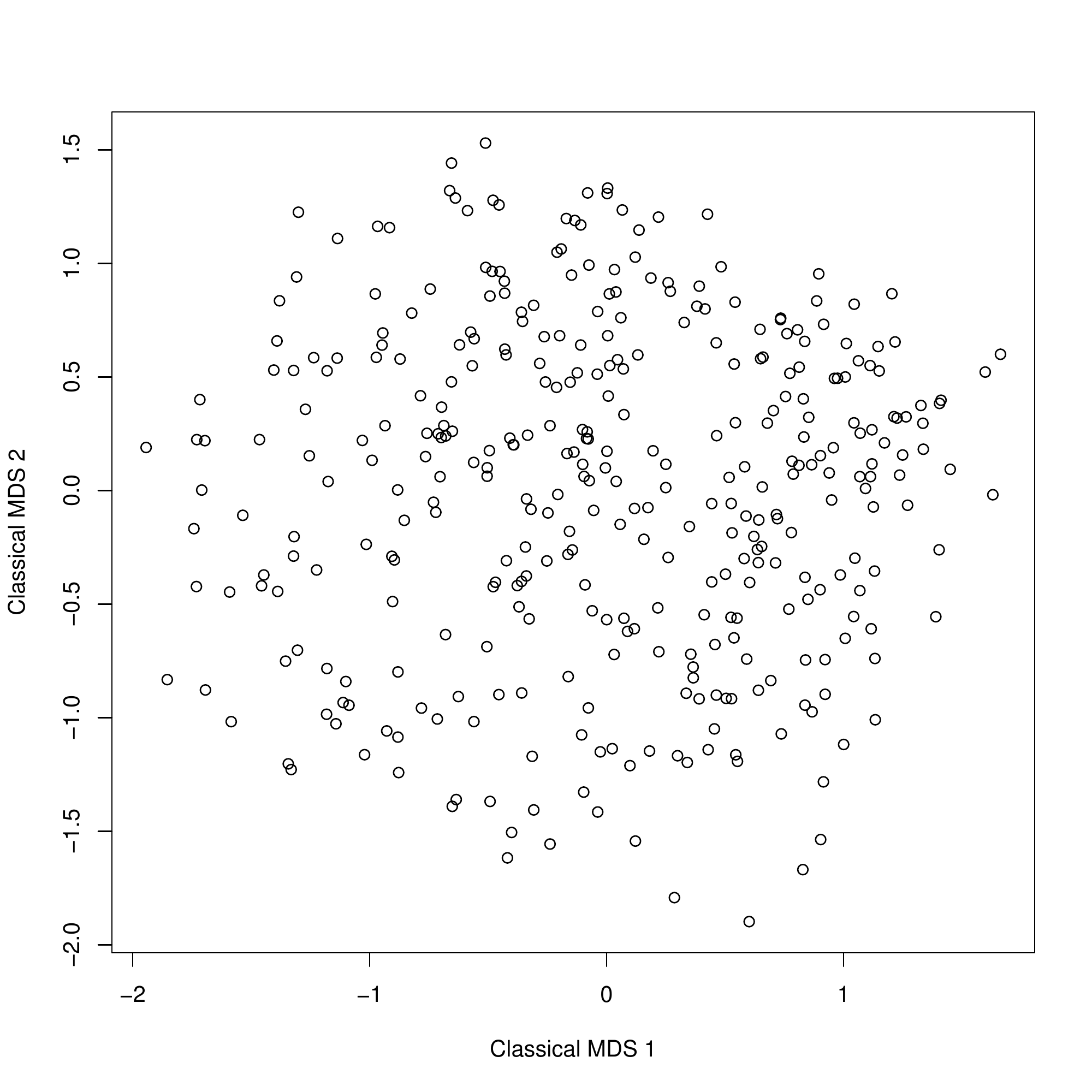}
\caption{2-dimensional classical multidimensional scaling for brand preferences. Left side Study 1 (London), right side Study 2 (Scotland).} \label{fmds}
\end{figure}

\begin{figure}[t]
\centering
\includegraphics[width=0.48\textwidth]{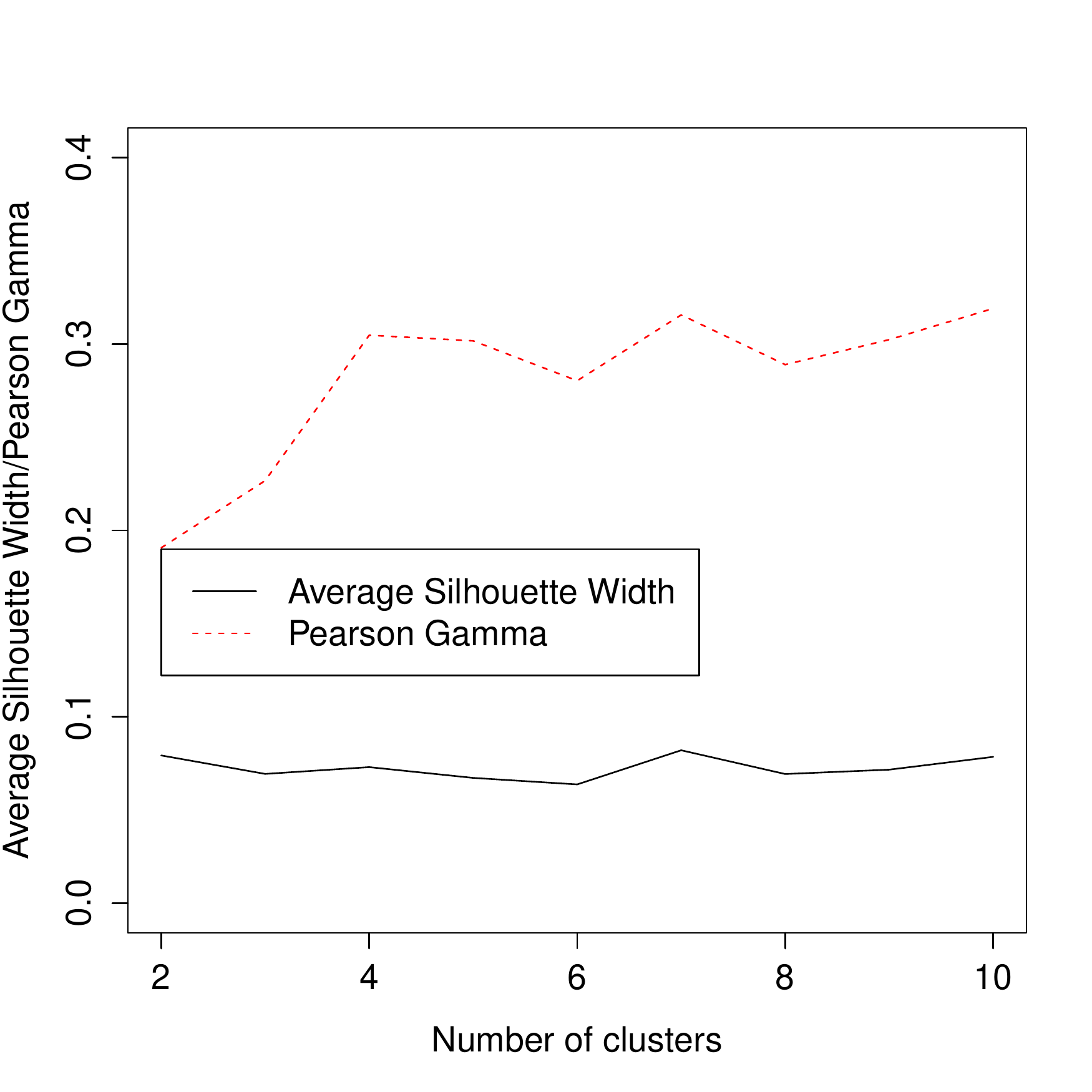}
\includegraphics[width=0.48\textwidth]{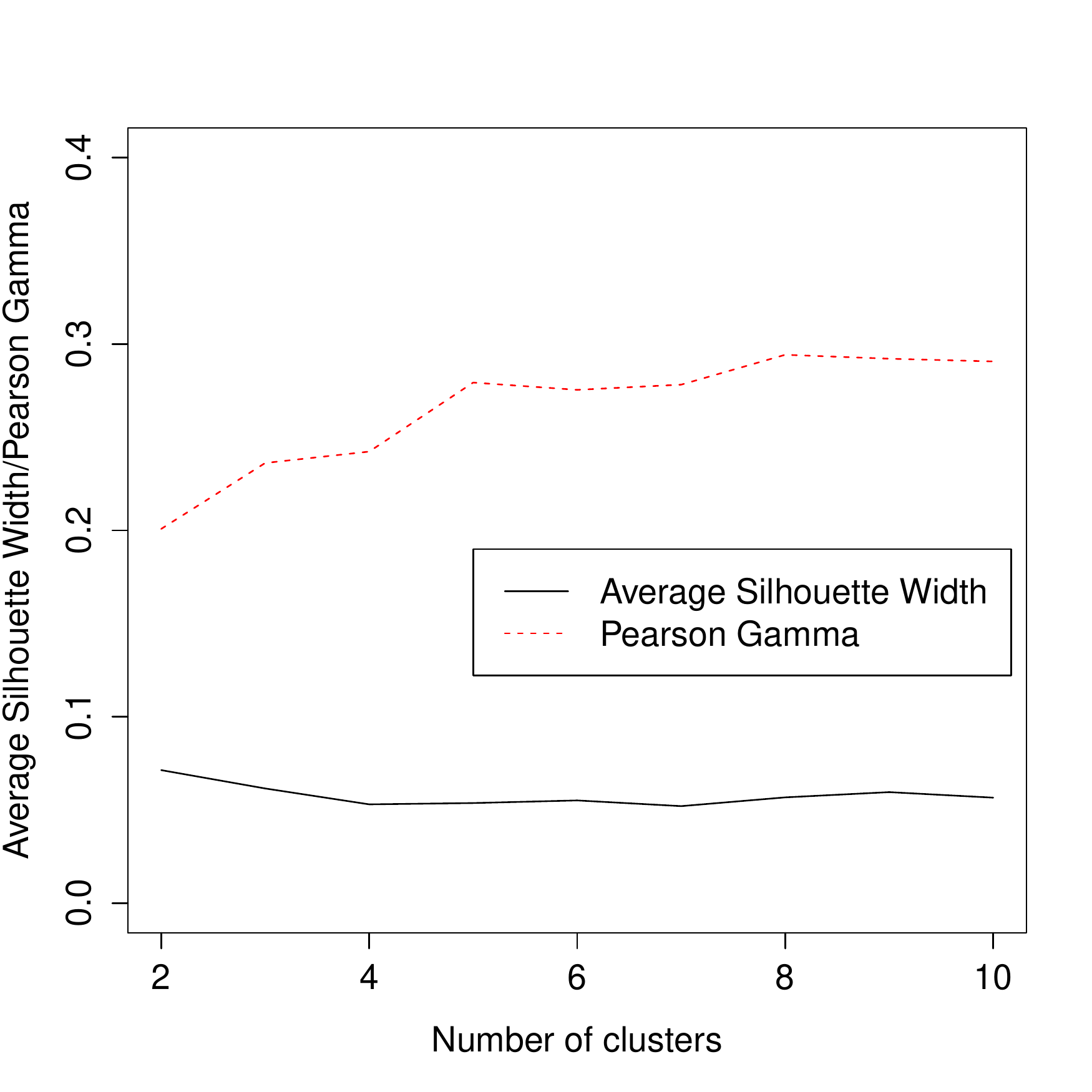}
\caption{ASW and P$\Gamma$ for PAM with 2-10 clusters. Left side Study 1 (London), right side Study 2 (Scotland).} \label{faswpg}
\end{figure}
We present the results using (\ref{e:sfootrule}) for all five brand categories combined in detail; detailed results for the individual categories can be obtained from the authors.

Figure \ref{fmds} shows MDS visualisations of the distances. The data from Study 1 do not show strong clustering, but two slightly denser areas can be observed, and on the left side some participants with rather atypical preferences are scattered. The data from Study 2 lack any visible clustering structure, although one can obviously still partition them into clusters with as low as possible within-cluster distances. A marketing practitioner could be interested in the specific clusterings, but we do not focus on them here. 

Figure \ref{faswpg} shows that the values of the ASW and P$\Gamma$ are rather low for both datasets (both of these are between -1 and 1 and values should be substantially higher than 0 in order to indicate clear clusters) but slightly better for Study 1. The plot for Study 1 can be seen as some weak indication in favour of $G=7$, but is still quite ambiguous when it comes to the optimal number of clusters; in Study 1, in agreement with the MDS, no specific $G$ looks convincing.

The five different categories of brands are rather heterogeneous; correlation coefficients between the vectors of distances from (\ref{e:sfootrule}) applied to individual categories are mostly between 0 and 0.1 with only three correlations (all in Study 1) between 0.1 and 0.2. All of these are correlated at around 0.4-0.5 with the distance based on all categories combined, which therefore seems to be a well balanced compromise. The adjusted Rand indexes \cite{hubertarabie1985rand} between the clusterings based on individual categories and combined categories are mostly around 0.1 (between $-0.017$ and 0.299), indicating a rather moderate similarity.

\begin{figure}[t]
\centering
\includegraphics[width=0.48\textwidth]{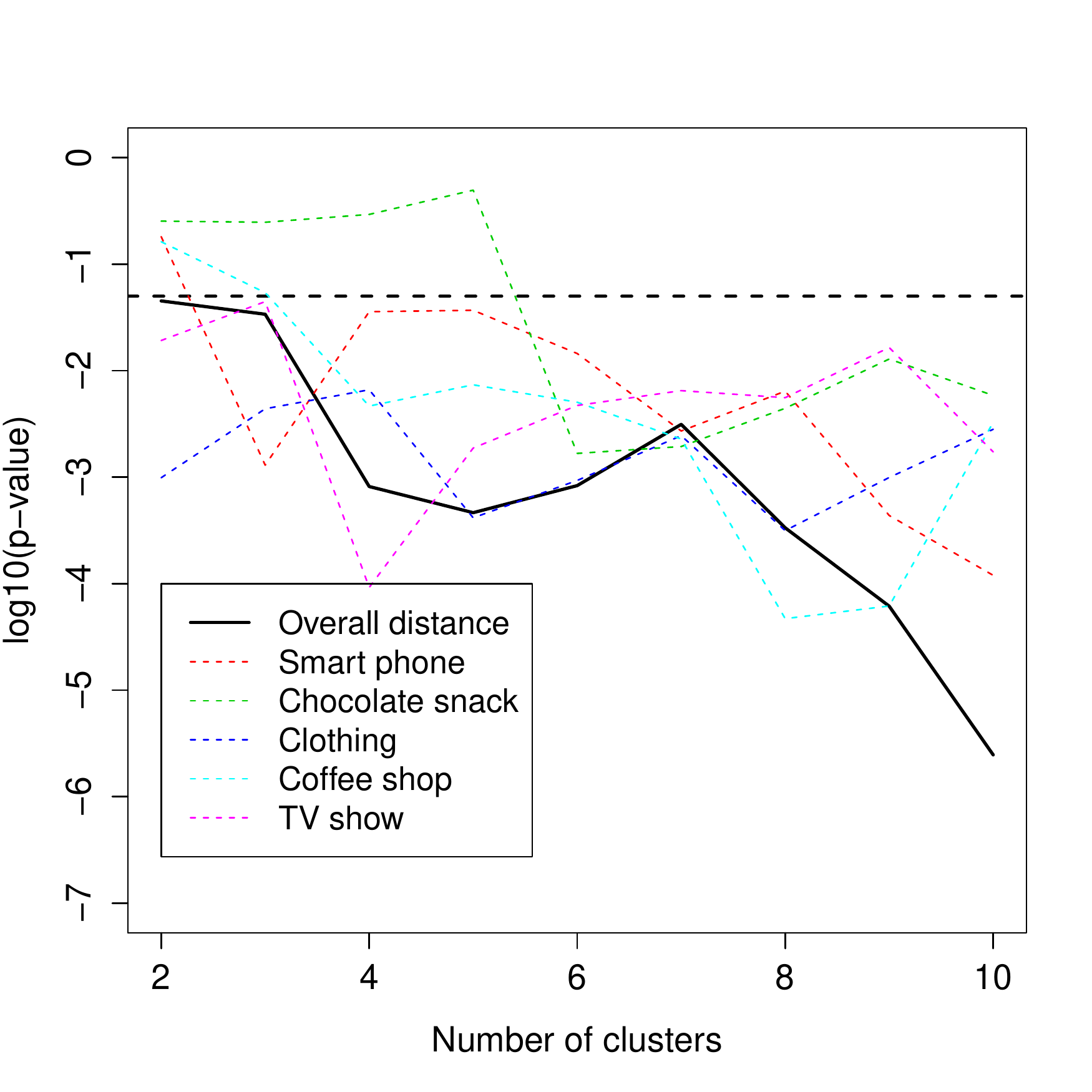}
\includegraphics[width=0.48\textwidth]{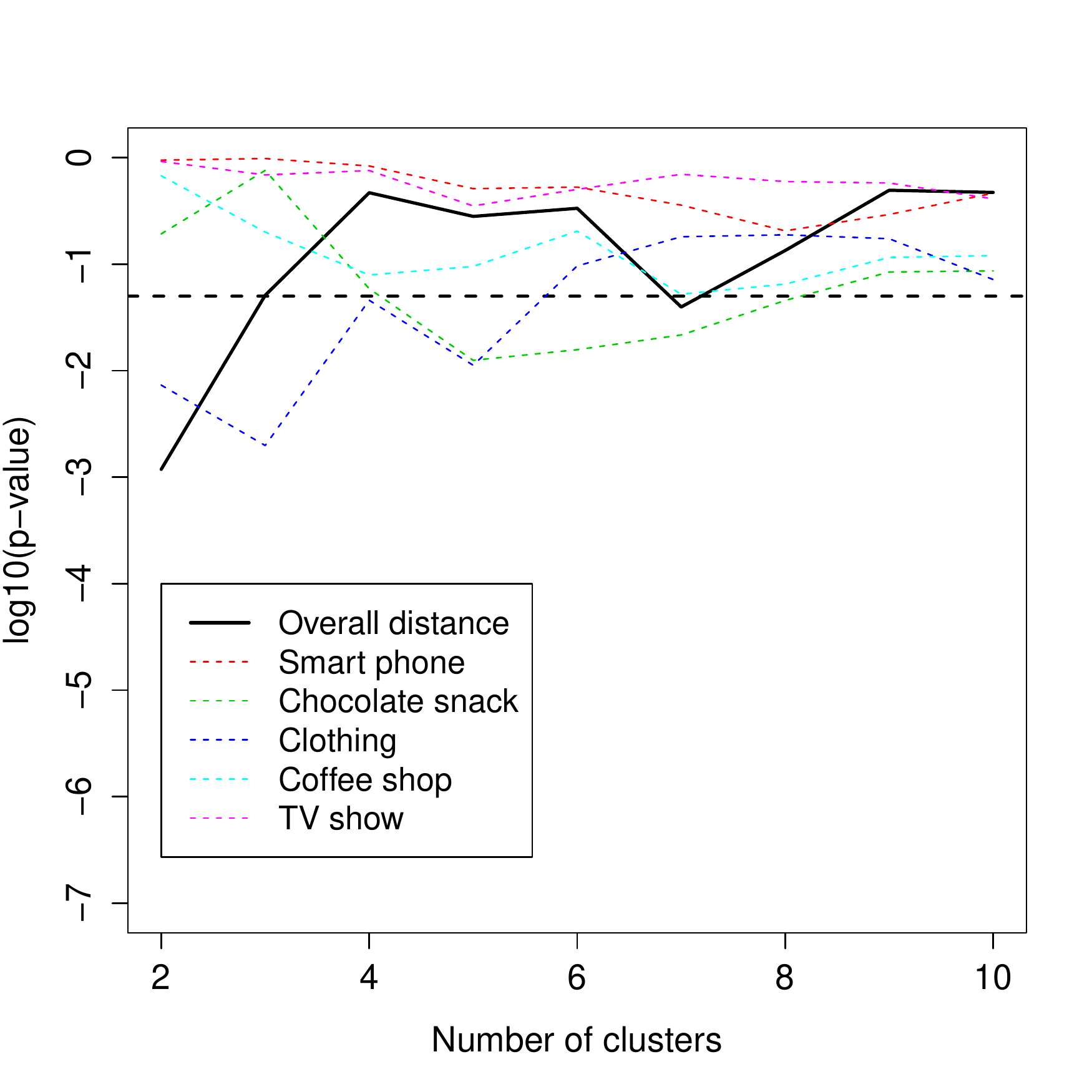}
\caption{log10-transformed p-values for testing the $H_0$ that all of the coefficients for the personality variables are zero from multinomial logistic regression with clustering with 2-10 clusters as response (the solid line corresponds to the clustering with all categories combined). Left side Study 1 (London), right side Study 2 (Scotland).} \label{fmultinomp}
\end{figure}

\begin{figure}[t]
\centering
\includegraphics[width=0.48\textwidth]{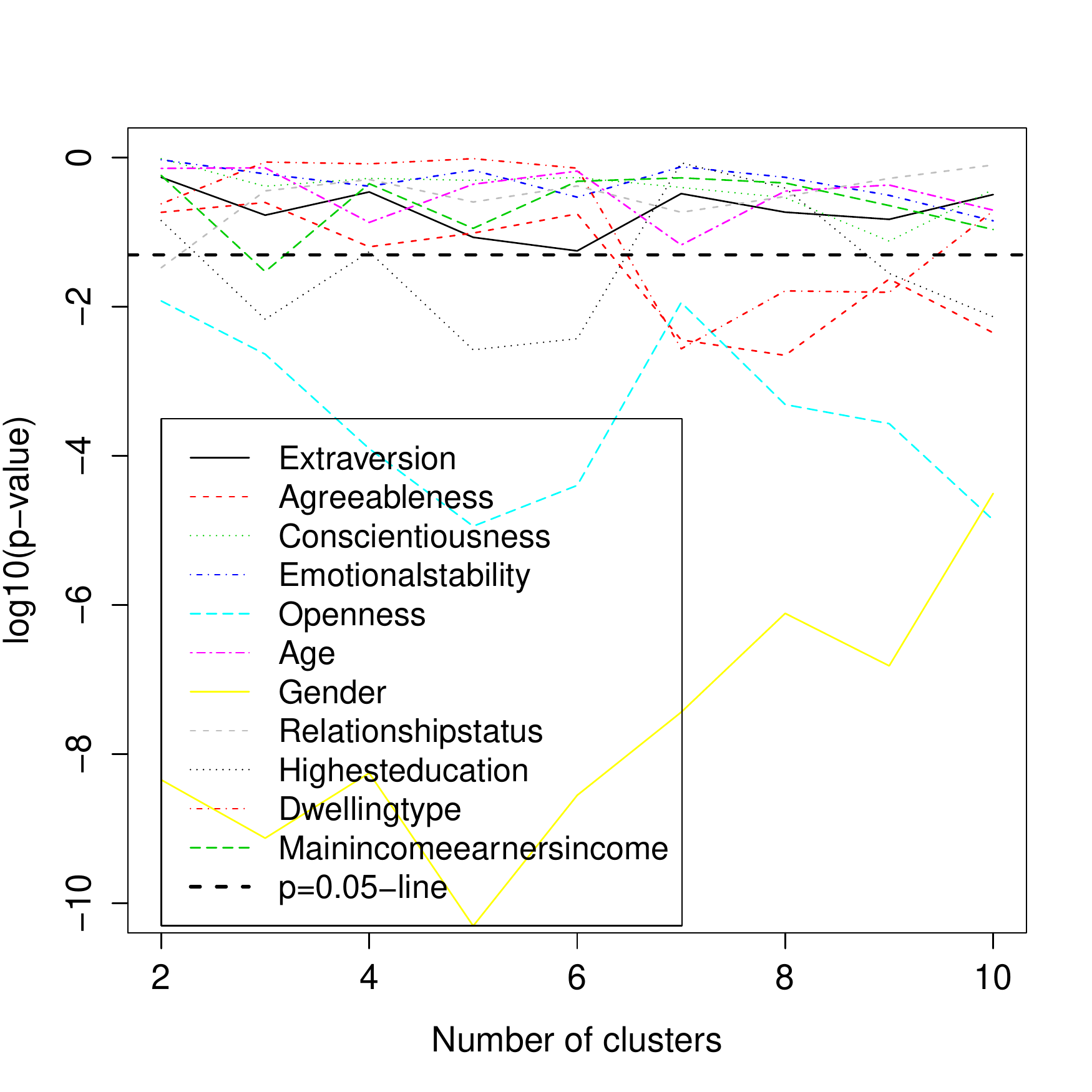}
\includegraphics[width=0.48\textwidth]{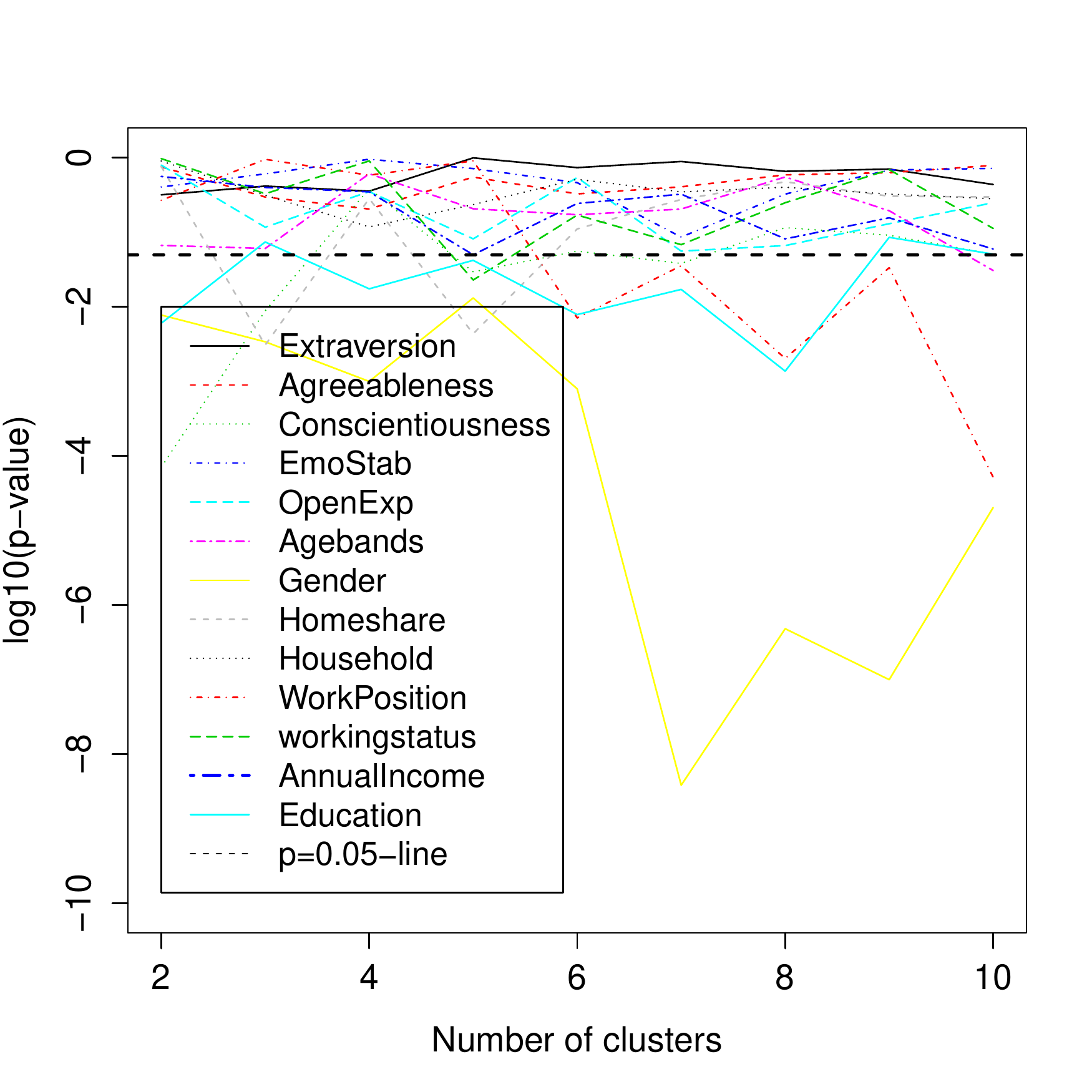}
\caption{log10-transformed p-values for testing the $H_0$ that the coefficients for every individual variable are zero from multinomial logistic regression with clustering with 2-10 clusters as response (the solid line corresponds to the clustering with all categories combined). Left side Study 1 (London), right side Study 2 (Scotland).} \label{fvarwise}
\end{figure}
Figure \ref{fmultinomp} shows the log10-transformed p-values for the inclusion of all five personality variables as one block into the logistic regression model in addition to the 6 socio-demographic variables across all nine cluster solutions. For Study 1, it shows that the contribution of the five personality variables is significant for all models at a significance level of 0.05 for all categories combined. Their weakest contribution is found for $G=2$; for $G\ge 4$ and particularly $G\ge 8$, the personality variables are strongly significant. Finer segmentation solutions show a clearer impact of the personality variables. The results for individual categories are in line with this.

For Study 2, Figure \ref{fmultinomp} shows that the personality variables taken together as a block, for all categories combined, only have a significant impact on predicting cluster membership in the multinomial logistic regression models for the 2-cluster solution. Given the number of tests, this is a very weak indication regarding the impact of the personality variables. Results for individual categories show some weakly significant impact for clothing and chocolate snacks.

Looking at the individual variables, Figure \ref{fvarwise} shows a breakdown of the association between each individual socio-demographic and personality variable and cluster membership for all categories combined. Gender and Openness to Experience are the only variables with significant associations across all cluster solutions in Study 1. In addition, Agreeableness has a significant influence for the more finer-grained solution from 7 to 10 clusters and Dwelling Type is also significant for the solutions with 7 to 9 clusters. The only other variable that shows a significant influence for at least some solutions is the Highest Education Level achieved. 

In Study 2, Gender shows a strongly significant influence over all clustering solutions. Homeshare, Education and WorkPosition are significant for some numbers of clusters. Out of the personality variables, Conscientiousness is significant 
for $G=2$ and $G=3$ though not for larger $G$. The corresponding results for the individual categories are summarized in Table \ref{tresults}.

\begin{figure}[t]
\centering
\includegraphics[width=0.48\textwidth]{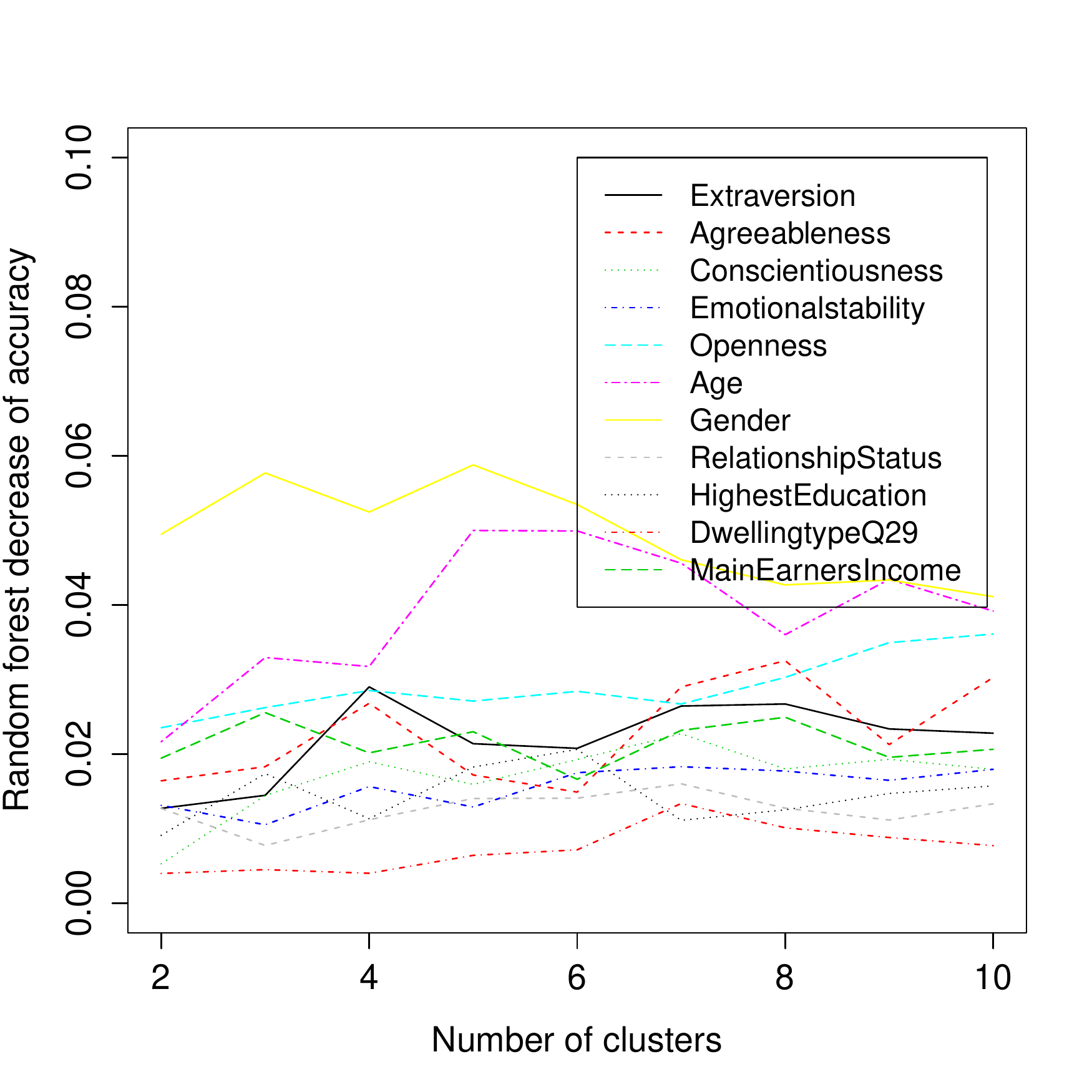}
\includegraphics[width=0.48\textwidth]{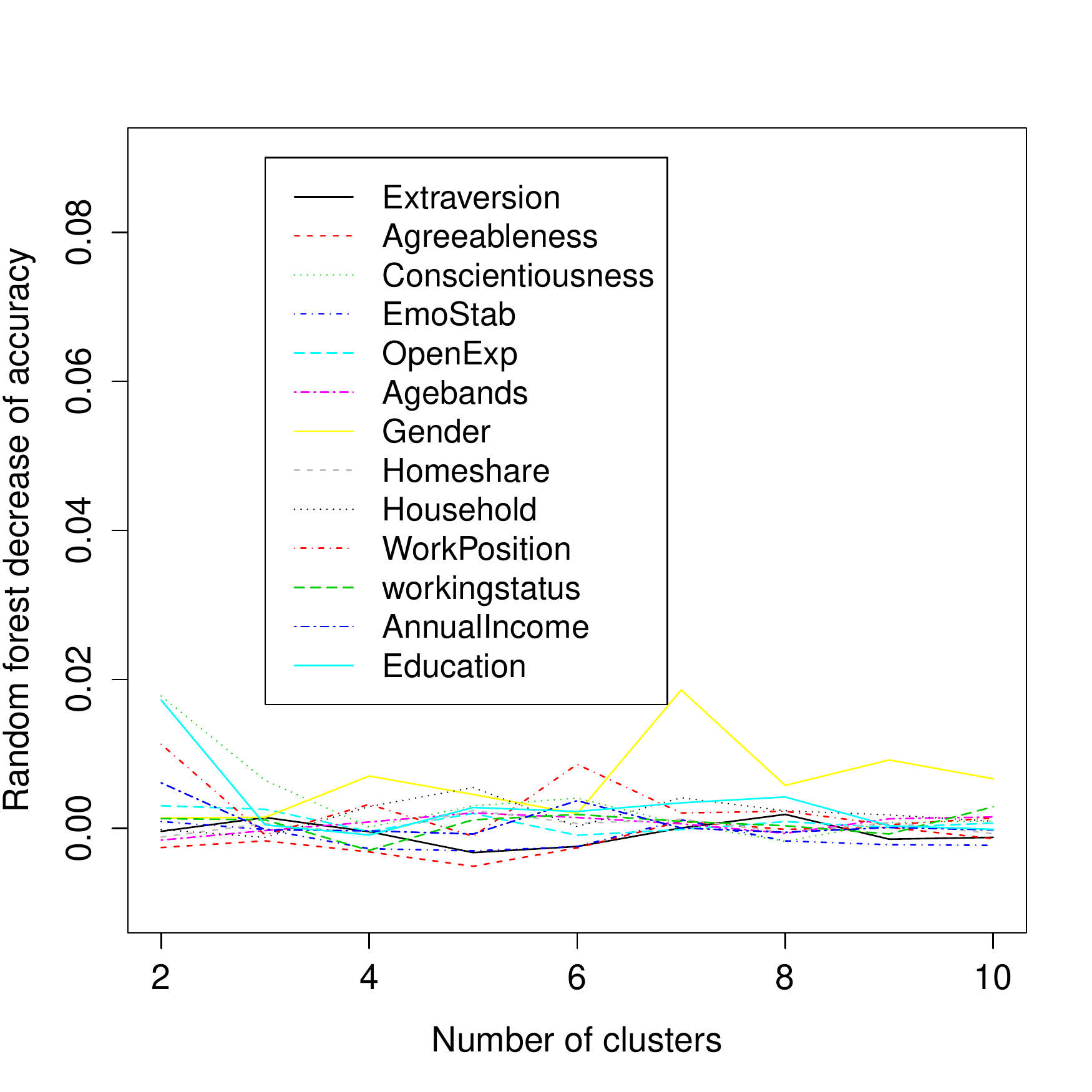}
\caption{Random forest decrease of accuracy for all variables. Left side Study 1 (London), right side Study 2 (Scotland).} \label{fforest}
\end{figure}

\begin{figure}[t]
\centering
\includegraphics[width=0.48\textwidth]{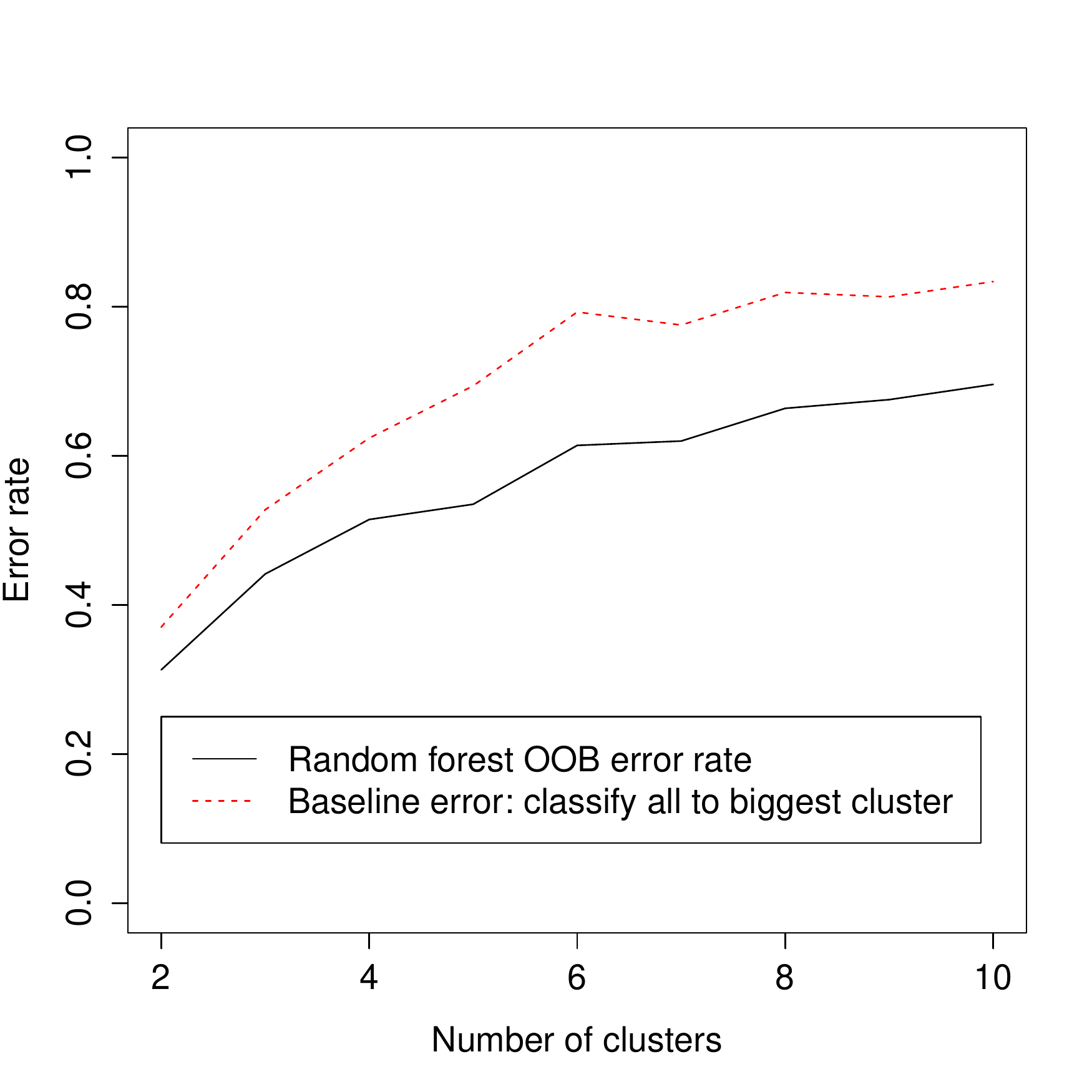}
\includegraphics[width=0.48\textwidth]{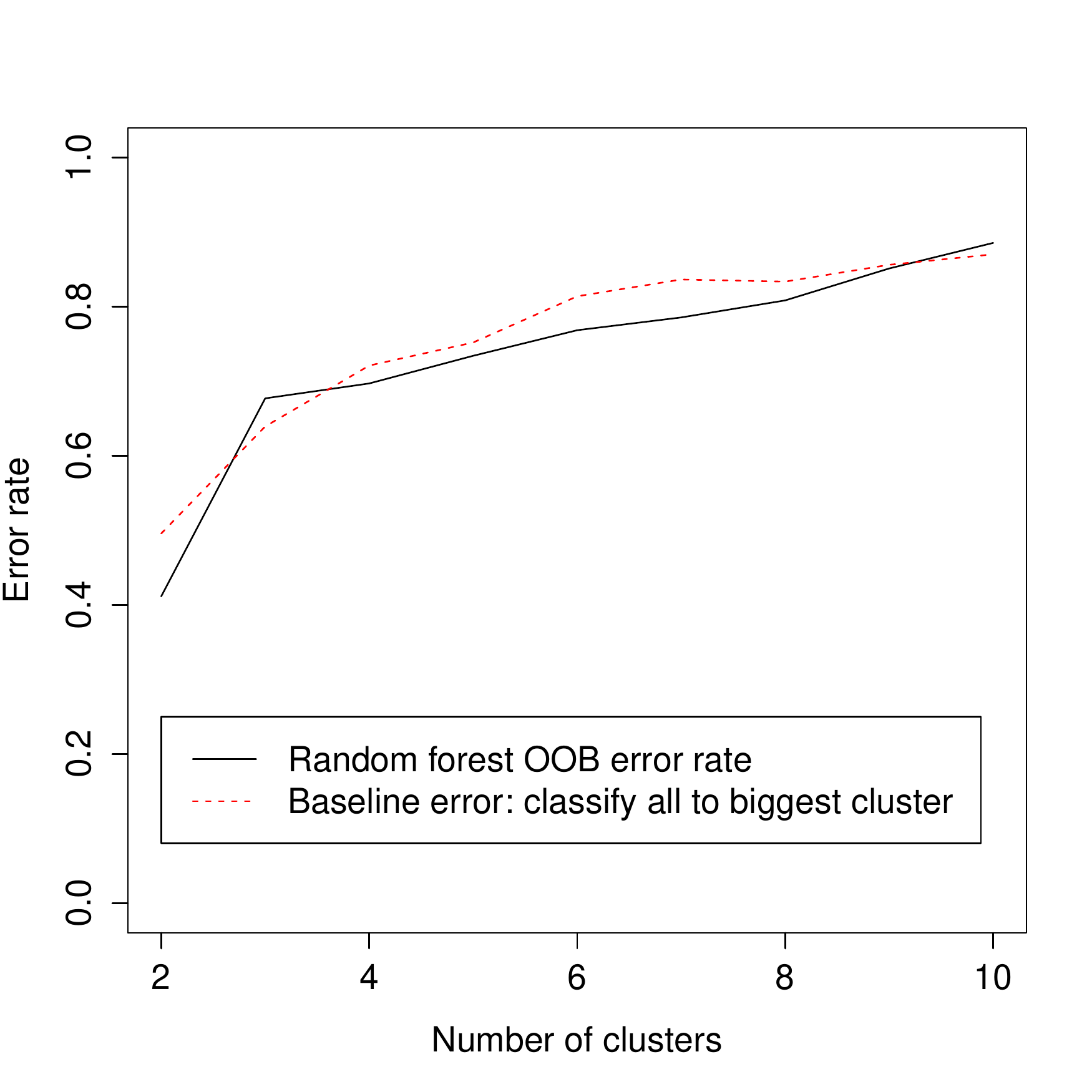}
\caption{Comparison of random forest OOB-error with error from the trivial rule to classify every observation into the largest cluster. Left side Study 1 (London), right side Study 2 (Scotland).} \label{fforesterror}
\end{figure}

The variable-wise results from the random forest model (Figure \ref{fforest}) in Study 1 largely confirm the findings from the logistic regression models regarding the importance of Gender and Openness to Experience. In addition, the trend for Agreeableness to gain importance for finer grained cluster solutions is also visible in the random forest models. However, in contrast to the results from the logistic regression models, Age plays an important role in the random forest models across all clusterings and Extraversion and Main Earner’s Income also rank among the more important variables for predicting cluster membership but without any clear trend in variable importance across cluster solutions. The fact that the importance of Age, Extraversion and Main Earner’s Income was not visible in the results given in Figure \ref{fvarwise} is due to correlations among some of the predictor variables and the different handling of the contribution of correlated predictors in the multinomial logistic regression compared to the random forest model. Specifically, Extraversion shows a considerable but not unusual (\cite{aluja2002comparative}) correlation of $r=0.42$ with Openness and in the current sample, Age was correlated with Gender ($r = -0.38$) and Main earner’s income ($r=-0.13$). While the specific contribution of each variable in the logistic regression model is assessed in addition to all other variables being present in the regression model, the random forest spreads the contribution more equally across correlated variables. From the present data it is not possible to decide whether the ‘true’ contribution to cluster membership does arise from Age or Gender or both. Therefore is seems sensible to consider both variables as potential contributors in line with the variable importance score from the random forest models. 

Figure \ref{fforesterror} depicts the OOB classification error of the random forest model across the nine cluster solutions and in comparison with the baseline error. We define baseline error as the error resulting from the assignment of all participants to the largest cluster. The random forest model consistently does a bit better but not much better than the baseline error in Study 1 (between 8\% for the 2-cluster solution and 18\% for the 6-cluster solution).

In Study 2 the random forest by and large hardly improves on the baseline error, which indicates that it does not provide specific evidence for the influence of any of the explanatory variables. Accordingly, Figure \ref{fforest} does not show clear differences between variables.

\begin{table}[t]
\begin{tabular}{lccc} \hline
  & MLR evidence for & MLR most & random forest most\\ 
  & personality variables & influential variables & influential variables \\ 
\hline 
\multicolumn{4}{l}{Study 1 London} \\
\hline 
Categories combined & strong & Gender, Open, & Gender, Age,  \\
& & Educ., Dwelling, & Open, Extraversion,\\
& & Agreeableness & Agreeableness \\
Smart phone & strong & Gender, Emotion, & Agree., Age \\
& & Conscientiousness & \\
Chocolate Snack & weak & Open, Relationship, & Age, Income \\
& & Gender & \\
Clothing & strong & Open, Education, & Age \\ 
& & Relationship, Income & \\
Coffee Shop & strong & Extraversion & Income, Age\\
& & Agreeableness & Extraversion \\
TV Show & strong & Gender, Emotion, & Emotion, Age, \\
& & Income, Open & Extrav., Gender\\
\hline 
\multicolumn{4}{l}{Study 2 Scotland} \\
\hline 
Categories combined & weak & Age, Education, & N/A\textsuperscript{a}  \\
& & WorkP., Homes. & \\
& & Conscientiousness &  \\
Smart phone & none & none & N/A\textsuperscript{a} \\
Chocolate Snack & weak & Open, Gender & N/A\textsuperscript{a} \\
Clothing & weak & Open, Age & N/A\textsuperscript{a} \\ 
Coffee Shop & none & none & N/A\textsuperscript{a}\\
TV Show & none & none & N/A\textsuperscript{a} \\ \hline
\end{tabular}
\caption{Overview of results (based on 
interpretation of the graphical displays). Note (a): Random forest not clearly better than baseline.}
\label{tresults}
\end{table}

\section{Discussion}\label{sdiscussion}
Our empirical results derived from the data of a sample of young adults living in a metropolitan area (Study 1) show that personality variables become more important for market segmentation as the number of target segments increases. Thus, personality variables play an important role in finer grained market segmentations which supports earlier evidence along the same lines \cite{choo2004type,mulyanegara2009big,ferguson2011personality}. However, the relationship between the systematic increase in the number of consumer groups  and the importance gained by the big five personality variables becomes very apparent  through the clustering empirical approach used here. The variable importance scores of the random forest model clearly indicate the primary importance of the socio-demographic variables, Age and Gender, but personality variables Openness, Extraversion and Agreeableness play a considerable role as well, at least segmentations with 6 or more target groups.

Hence, the data from this study does not agree with Sharp’s notion that brand preferences are not at all linked to consumer traits.  However, our methodological procedure differed from previous studies (e.g. the widely cited study by Evans \cite{evans1959psychological}) in as much as we aggregated brand preferences across five product categories to cluster consumers (as opposed to predicting preferences for a single brand). Thus, this study used a different dependent variable compared to the earlier studies cited by \cite{sharp2010brands}. Moreover, for this study we selected a specific sample of young adults living in an urban area and matched product categories and brands to it that are most relevant and most familiar to this narrow demographic group. Thus, in marketing terms the sample represents a main target audience for the selected brands. 

Study 2 uses the same brands and survey set-up but presents them to a sample of adults from a different geographical area in Britain with much more heterogeneous demographic characteristics.  
The demographically much broader sample in Study 2 does not yield such positive results. This sample was not only much more heterogeneous in its socio-demographic characteristics but also less representative of the chosen brands target audience which might have resulted in many more random preference ratings. 

So overall there are mixed results regarding whether personality variables can have predictive power for distinguishing between groups of consumers, which can probably explained by the different characteristics of the samples regarding homogeneity and representation of the target audience of the brands, although this is somewhat speculative given that there are only two samples. In any case the results from the two samples in the present study already indicate that the association between personality and demographic variables on one hand and consumer cluster of brand preferences on the other hand is not universal but does depend on the sample of consumers.

From the present data we could not replicate the result reported by \cite{sandy2013predicting} that personality variables are generally as important for market segmentation as demographic variables. But at least for the data in Study 1 our results demonstrate that personality variables can have a considerable impact on market segmentation solutions in combination with socio-demographic variables. However, it is worth remembering that in the present study we used a 10-item personality short form to measure traits on five personality dimensions while \cite{sandy2013predicting} used four dimensions of the much longer NEO- PI-R inventory. While the pseudo $R^2$-values reported by \cite{sandy2013predicting} are not directly comparable to the classification rates and variable importance scores obtained from our random forest model, it appears nonetheless that the measures derived from 10-item short form had a similar relative impact for the finer-grained segmentation solutions in Study 1.

In Study 1 the classification model performed clearly better than the baseline error rate, refuting the notion that market segmentation based on consumer background variables is a largely impossible undertaking (\cite{sharp2010brands}). However, one has to bear in mind that we used a different dependent variable compared to the earlier studies that \cite{sharp2010brands} and others have referred to as evidence for the ineffectiveness of market segmentation. Hence, it is possible that Sharp’s interpretation and the results from our clustering approach are ultimately compatible if the aggregate of several brand preferences represents more stable measurement than single brand preferences. Additionally, the use of the random forests as a modern data mining technique with a strong predictive power might also have contributed to the clear segmentation results of this study.  

In sum, the results of this study have shown that is possible to segment consumers according to their brand preferences into clusters. While there seemed to be no ‘optimal’ or ‘natural’ number of clusters, any division of the sample in Study 1 into clusters derived from the PAM clustering procedure had a significant association with socio-demographic and personality variables, with personality variables being more important in segmentation solutions with a higher number of consumer groups. This means for marketing practitioners that the decision on the number of target segments of a market can be based on practical considerations and costs. The results from Study 1 show that market segmentations with a classification accuracy of up to 70\% can be achieved with as few as 11 variables derived from 16 question items. Given that several variables (e.g. Emotional Stability/Neuroticism, Conscientiousness, Dwelling Type) did not contribute much to the prediction accuracy of the random forest models, it seems possible to reduce the length of the consumer questionnaire even further without losing a large amount of predictive power.
 
In addition, the segmentation approach we propose here has the conceptual advantage that it generates data-driven model solutions that are very close to practical notions of brand personality (\cite{aaker1997dimensions}) and profiles of prototypical consumers which are often defined by associations between habits, preferences and demographic profiles and which often seem helpful to guide the development of creative ideas and planning of marketing campaigns. Thus, this clustering approach to market segmentation which builds on personality and socio-demographic variables not only has comparatively good predictive power but can also blend in conceptually with current practice in advertising and marketing.

On the statistical side, a key feature of this paper is that our use of clustering does not rely on the existence of a true underlying clustering or a true number of clusters. Clustering is used to summarize and simplify the brand preferences, and analyses run over a range of numbers of clusters rather than a single one. Because the clusterings are constructed so that large within-cluster distances are avoided, clusters can still be interpreted by marketing practitioners, while their ``constructive'' rather than ``real'' nature is acknowledged. Another observation is that both the multinomial logistic regression and the random forest deliver valuable and largely complementary information about the impact of the various variables on the clusterings.

The methodology can be used by a practitioner to use and interpret specific clusters. The estimated coefficients of the multinomial logistic regression can be interpreted regarding the specific contribution of the variables regarding specific clusters, although this is beyond the scope of this paper.

A very elegant option for future studies as suggested by an anonymous reviewer would be to combine clustering and the fit of explanatory models for the clusters into a single model. However, this beyond the scope of the present paper work and standard tests of the MLR could no longer be used within such an approach.

\section*{Funding and Acknowledgements}
The work of Christian Hennig was supported by EPSRC grant EP/K033972/1. We thank Clare Prakel and Les Binet from adam\&eveDDB UK for their help with the data from the TGI database and general advice and support throughout of this project. We also thank Anne MacLean for the data collection through the ScotPulse panel.

\bibliographystyle{tfs}
\bibliography{MarketSeg_References}

\end{document}